\documentclass[a4paper,11pt]{article}
\usepackage{pos}

\usepackage{lineno}
\title{Energy dependence of light neutral meson $p_{\rm T}$ spectrum produced in pp collisions at the LHC measured in ALICE}
\ShortTitle{Light neutral mesons in pp collisions}

\manuallySeparateAuthors
\author*{Pooja Pareek}
\author{ for the ALICE collaboration}

\affiliation{Variable Energy Cyclotron Centre,\\
  1/AF, Bidhannagar, Kolkata-700064, India}

\emailAdd{pooja.pareek@cern.ch}
\abstract{We present the measurement of light neutral mesons, $\pi^{0}$ and $\eta$, in pp collisions at different center-of-mass energies obtained with the ALICE experiment at the LHC. 
The $\pi^{0}$ and $\eta$ mesons are measured via photons reconstructed by the electromagnetic calorimeters and the central tracking system.
The invariant cross-section of $\pi^{0}$ and $\eta$ mesons are measured in a broad $p_{\rm T}$ range at $\sqrt{s} = 0.9, 2.76, 7, 5.02$ and 8 TeV. The spectra of $\pi^{0}$ and $\eta$ mesons measured in pp collisions at different collision energies show $x_{\rm T}$-scaling at high $p_{\rm T}$ and violation of $m_{\rm T}$-scaling at low $p_{\rm T}$.
The smaller $x_{\rm T}$-scaling exponents of our measurements compared to RHIC may hint at a reduced importance of higher twist processes at LHC.

}

\FullConference{%
  The Eighth Annual Conference on Large Hadron Collider Physics-LHCP2020 \\
  25-30 May, 2020\\
  online}

\begin{document}
\maketitle

\section{Motivation}

The measurement of inclusive hadron production in pp collisions is a widely used tool to validate perturbative quantum chromodynamics (pQCD). The cross section of hadron production at high transverse momentum in pp collisions can be calculated within pQCD as these are produced by fragmentation of quarks and gluons from hard parton scatterings. 
Therefore, the study of light neutral mesons,  $\pi^{0}$ and $\eta$ in pp collisions in a wide transverse momentum range is important to understand parton dynamics by probing parton distribution functions and fragmentation functions in a new energy domain. In addition, $\pi^{0}$ and $\eta$ meson transverse momentum measurements are necessary to estimate the decay photon background for direct photon measurements and also to provide a reference for heavy-ion collisions. 
We present measurement of $\pi^{0}$ and $\eta$ mesons in pp collisions at LHC energies $\sqrt{s} = 0.9, 2.76, 7, 5.02$ and 8 TeV with the ALICE experiment.

\section{Neutral meson measurement}

The $\pi^{0}$ and $\eta$ mesons are reconstructed by invariant mass analysis of their decay into two photons. In ALICE the decay photons are measured with different methods. 
Photons can be measured in electromagnetic calorimeters: EMCal \cite{EMC} and PHOS \cite{PHOS}, designed to measure energy and hit coordinates of photons and electrons. 
Additionally, Photon Conversion Method (PCM) \cite{[1]} which is based on photon conversion in detector material ($\gamma \rightarrow e^+e^-$) is a complementary method to measure neutral mesons via conversion photons up to very low $p_{\rm T}$ ($\sim$ 0.3 GeV$/c$) using the central tracking system in ALICE. The energy resolution of the calorimeters improves with increasing $p_{\rm T}$ which together with the calorimeter trigger allows to measure neutral mesons up to high $p_{\rm T}$. In addition, a hybrid method is also used where one photon is reconstructed in a calorimeter and the second photon with the PCM method \cite{[2]}. Results from different techniques are combined to provide precision measurements of $\pi^{0}$ and $\eta$ in pp collisions over much wider $p_{\rm T}$ range than any other identified hadrons.

\section{Results}
\begin{figure}
\begin{center}

%left- 1920:2172  right-1920:3022 for ratio
\includegraphics[width=79.55mm,height=90mm]{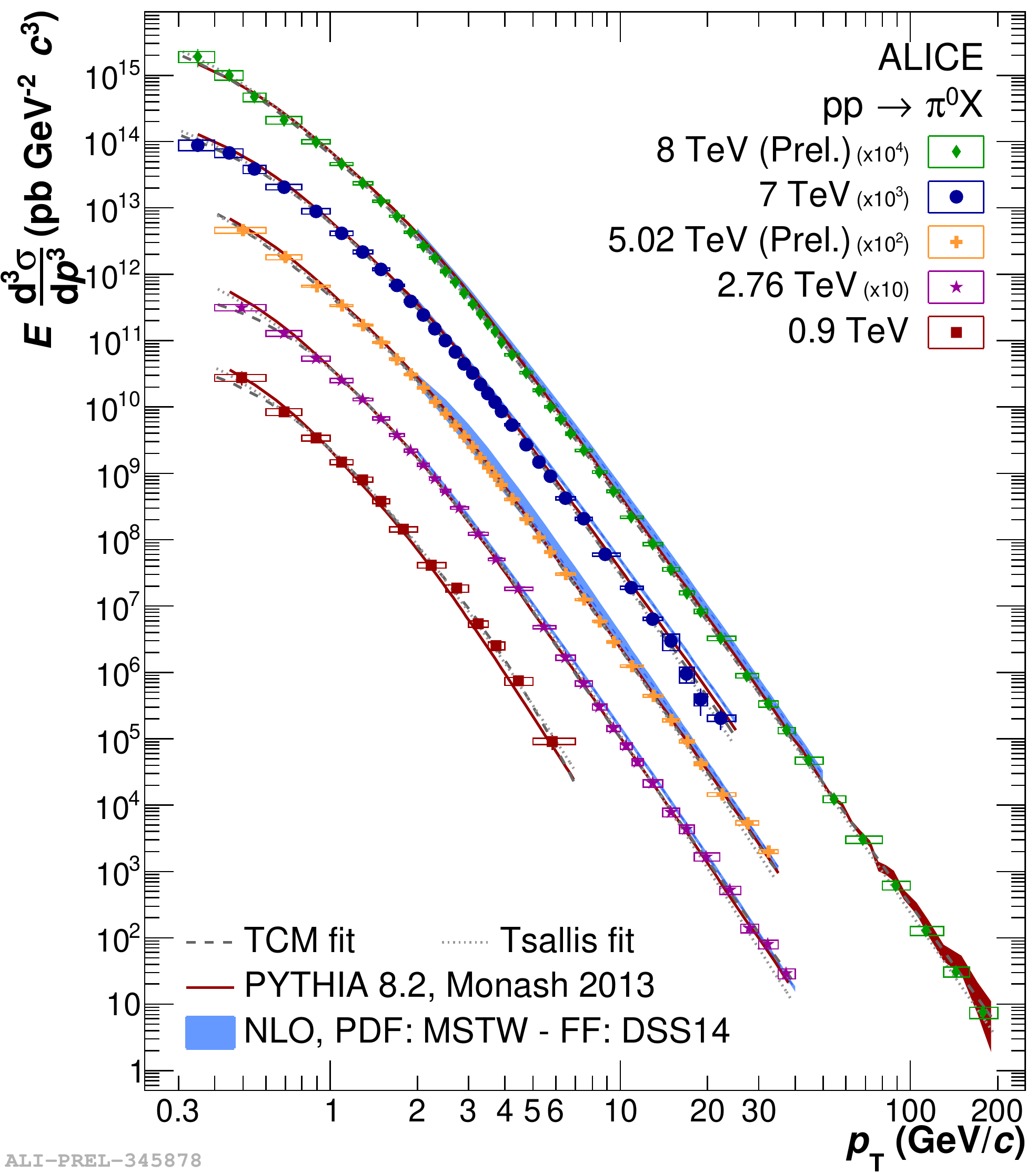}
\includegraphics[width=57mm,height=90mm]{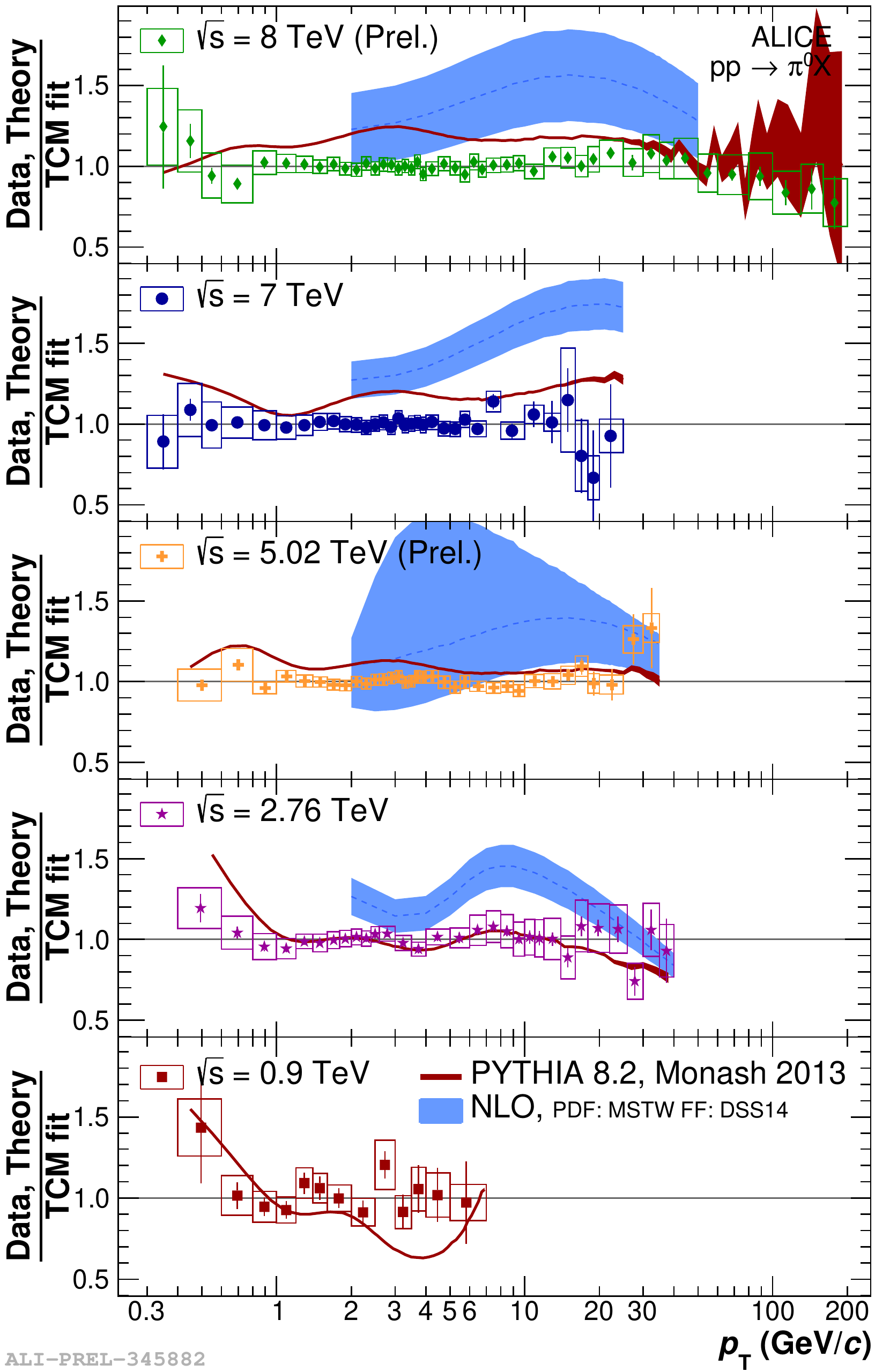}

\caption{\label{fig1} (Left) Invariant cross section of $\pi^{0}$ in pp collisions at $\sqrt{s} = 0.9, 2.76, 7, 5.02$ and 8 TeV \cite{[1],[2],[3]}. Neutral pions are measured in the $p_{\rm T}$ range $0.3 - 200$ GeV$/c$. (Right) The experimental data are compared to NLO pQCD calculation and PYTHIA. The ratio of data and theory predictions to TCM fit are shown in seperate panels for different energies.}
\end{center}
\end{figure}

%\subsection{Results}
In ALICE, $\pi^{0}$ and $\eta$ mesons are measured in pp collisions at $\sqrt{s} = 0.9, 2.76, 7, 5.02$ and 8 TeV \cite{[1],[2],[3]} as shown in Fig.\ref{fig1}. The results are described by empirical parametrizations like Two Component Model (TCM) \cite{tcm} and Tsallis function \cite{tsallis}. Further, invariant cross sections are compared with Next-to-Leading Order (NLO) pQCD calculation of $\pi^{0}$ and $\eta$ meson production with latest fragmentation functions (DSS14 \cite{[4]}, AESSS \cite{[5]}). NLO calculation overestimates the data by 20-50\% for pions and nearly a factor 2 for eta mesons. PYTHIA 8.2 \cite{[6]} with Monash 2013 tune reproduces $\pi^{0}$ spectra at high $p_{\rm T}$ but shows a deviation from data at moderate $p_{\rm T}$ in higher energy. Likewise, PYTHIA 8.2 \cite{[6]} with Monash 2013 tune describes the $\eta$ meson spectra. Overall PYTHIA shows better agreement with data compared to NLO pQCD.

\subsection{$x_{\rm T}$ scaling}
The invariant cross-section can be expressed as a function of new variable, $x_{\rm{\it}T} = 2p_{\rm T}/\sqrt{s}$ \cite{xt},

\begin{equation}
E \frac{d^{3}\sigma}{dp^{3}} = \frac{1}{\sqrt{s}^{n}}G(x_{\rm{\it}T}).
\end{equation}

Thus, from the above equation, the invariant cross-section as a function of $x_{\rm T}$, when multiplied by $\sqrt{s}^{n}$ does not depend on collision energy, which is referred to as $x_{\rm T}$ scaling.

\begin{equation}
\sqrt{s}^{n} \times E \frac{d^{3}\sigma}{dp^{3}} = G(x_{\rm{\it}T}).
\end{equation}

The scaling parameter $n$ is deduced from comparison of $x_{\rm T}$ spectra at different center-of-mass energies:

\begin{equation}
\label{npar}
 n (x_{\rm T}, \sqrt{s_{1}}, \sqrt{s_{2}}) = \frac{ln(\sigma^{inv}(x_{\rm T}, \sqrt{s_{2}})/\sigma^{inv}(x_{\rm T}, \sqrt{s_{1}}))}{ln(\sqrt{s_{1}}/ \sqrt{s_{2}}) }
\end{equation}

The dependence of the parameters $n$ on $x_{\rm T}$ for $\pi^{0}$ at different ALICE energies are shown in Fig.~\ref{fig2}. The parameter $n$ depends on both $x_{\rm T}$ and $\sqrt{s}$ at low $p_{\rm T}$ and shows an asymptotic trend at $p_{\rm T}$ > 2 GeV/c. We fit $n$ at $p_{\rm T} > $2 GeV$/c$ with a constant function and obtain $ n (\pi^0) = 5.03 \pm 0.06$ and $n (\eta) = 4.79 \pm 0.09$. For jets and photons a value of $n \simeq 4.5$ was measured in previous experiments \cite{[8]}; At RHIC, a value of $n = 6.38$ for $\pi^{0}$ was obtained \cite{[7]}.
In pQCD, hadron production at large transverse momentum are described by the leading (LT) and higher twist (HT) processes \cite{[8]}. In LT processes, $n = 4$ whereas $n > 4$ for HT processes \cite{[8]}. 
The difference in $n$ at RHIC and LHC may be related to different mechanisms for hadron production at these energies. Smaller value of $n$ for $\pi^{0}$ production at the LHC energies with respect to the RHIC energies suggests that the contribution of the HT processes is reduced with the growth of the collision energy. 
The $x_{\rm T}$ scaled $\pi^{0}$ and $\eta$ meson cross-sections were checked with calculated $n$ values, and both $\pi^{0}$ and $\eta$ mesons follows the $x_{\rm T}$ scaling at $p_{\rm T}$ > 2 GeV/c. The deviation from scaling at low $x_{\rm T}$ is attributed to the contribution from soft processes at low $p_{\rm T}$.

\subsection{$m_{\rm T}$ scaling} 

\begin{figure}
\begin{center}

\includegraphics[width=75mm,height=48.7mm]{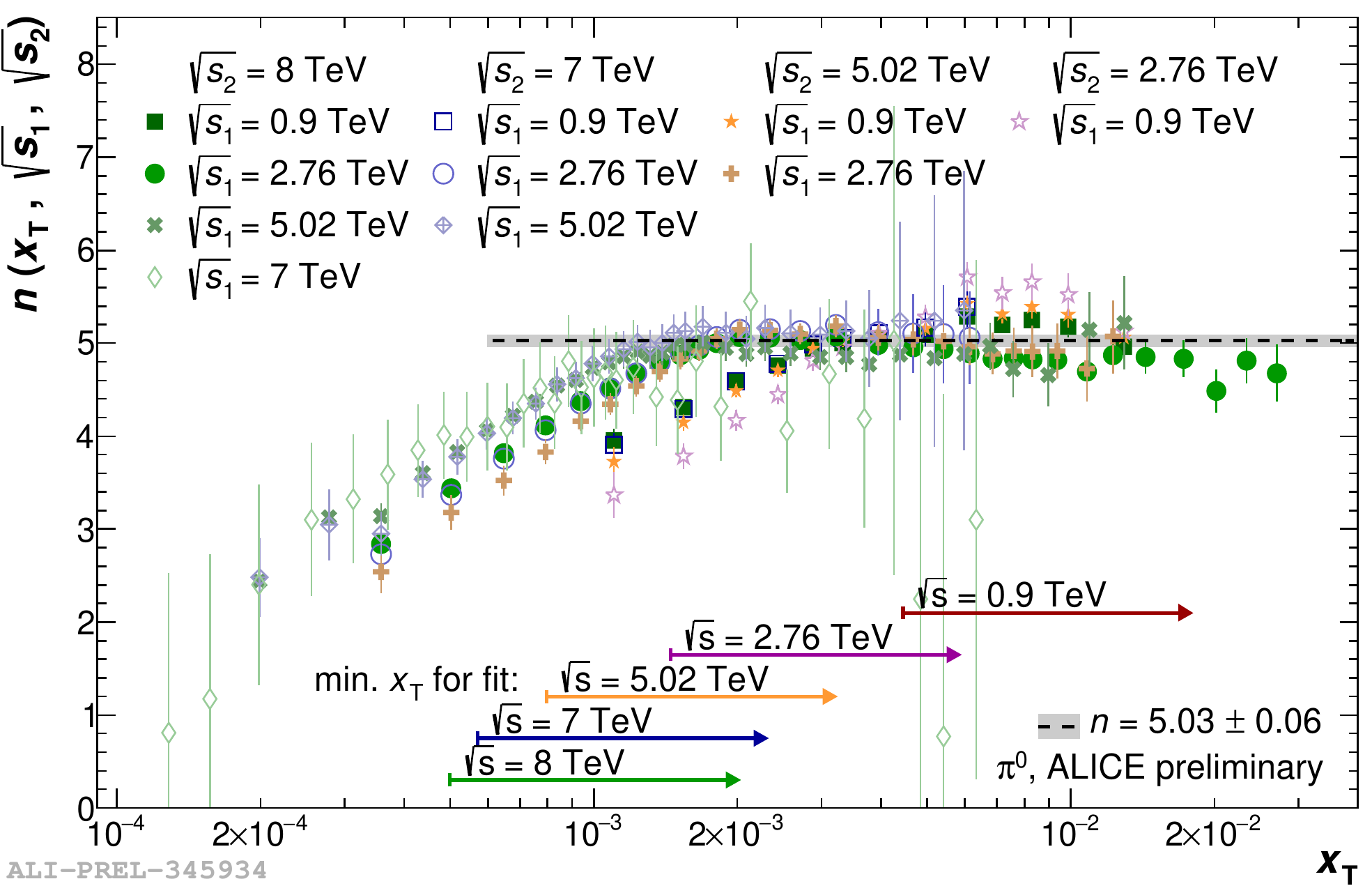}
\includegraphics[width=75mm,height=48.7mm]{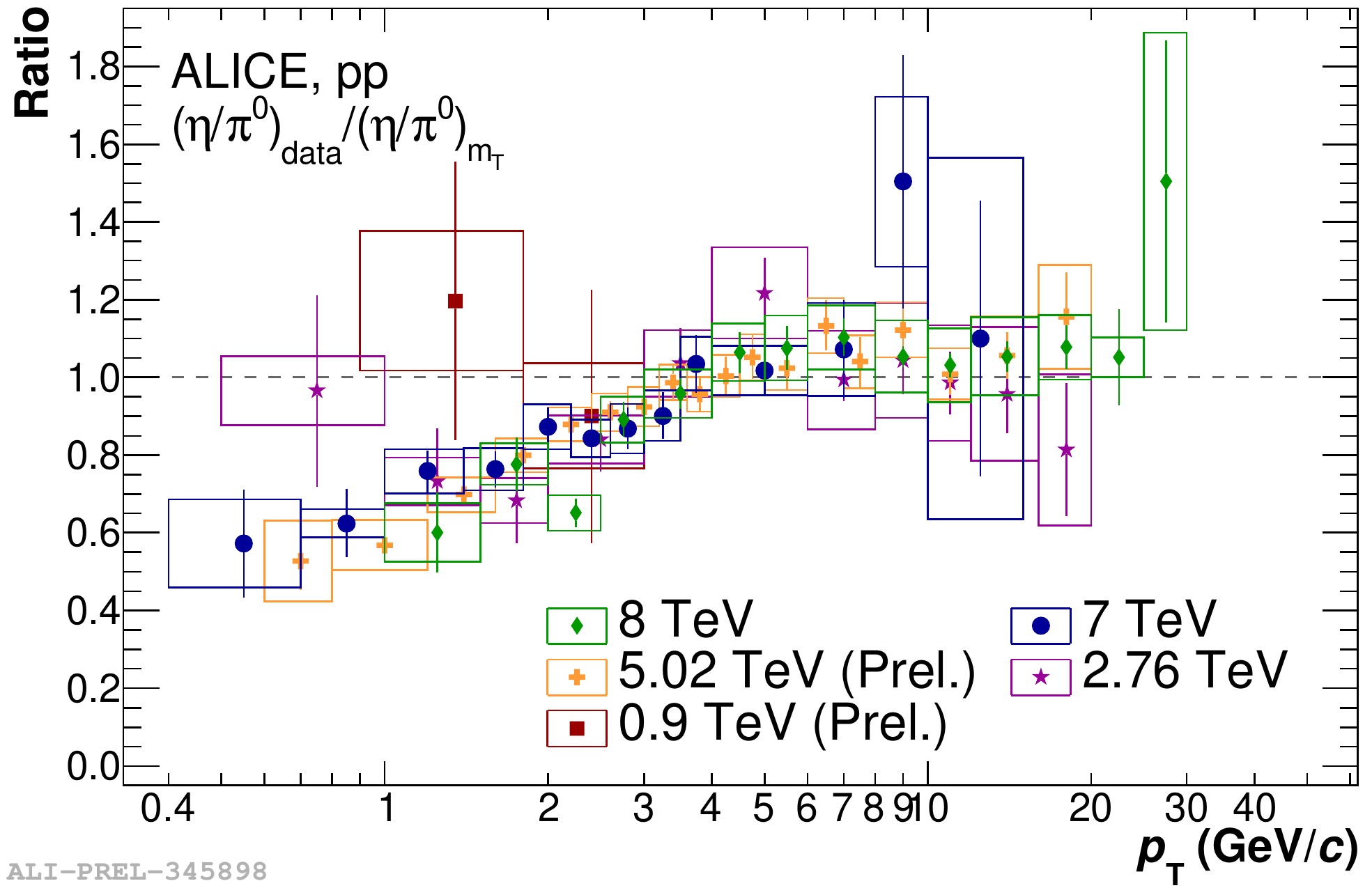}

\caption{\label{fig2} (Left) $n$ parameter for $\pi^{0}$ in pp collisions at different collision energies $\sqrt{s}$. (Right) Ratio of transverse momentum distributions of measured $\eta/\pi^{0}$ ratio over $\eta/\pi^{0}$ ratio obtained with $m_{\rm T}$ scaling for pp collision at  $\sqrt{s} = 0.9, 2.76, 7, 5.02$ and 8 TeV. }
\end{center}
\end{figure}

In lower energy experiments \cite{mT1,mT2,mT3,mT4}, spectral shapes of hadron species are found to be identical in shape when plotted as a function of $m_{\rm T}$ ($m_{\rm T} = \sqrt{p_{\rm T}^{2} + m^{2}}$), known as $m_{\rm T}$ scaling. $\eta/\pi^{0}$ ratio from data over $\eta/\pi^{0}$ ratio obtained with $m_{\rm T}$ scaling is shown in Fig.\ref{fig2}. We observe violation of the empirical $m_{\rm T}$ scaling at $p_{\rm T} <$ 2 GeV$/c$ at the LHC, which is important in tuning the phenomenological models at low $p_{\rm T}$.

%aspect ratio: left-1920:1229,  right-1920:1248 (pixel width same for all)

\section{Summary}
Measurements of neutral meson spectra, and $x_{\rm T}$ and $m_{\rm T}$ scaling have been performed with the ALICE detector. 
These measurement can be used to validate pQCD predictions. The measured spectra in pp collisions at $\sqrt{s} = 2.76, 7, 5.02$ and 8 TeV show disagreement with pQCD calculations. 
The $m_{\rm T}$ scaling of measured spectra show violation of this empirical scaling at $p_{\rm T} <$ 2 GeV$/c$. We study $x_{\rm T}$ scaling and observe the $n$ parameters for $\pi^{0}$ and $\eta$ are consistent with jets and direct photons, which suggests that almost all neutral mesons are generated by jet fragmentation at the LHC. In contrast at RHIC, neutral mesons are possibly generated by both jet fragmentation and direct hadron production. 
Therefore, $x_{\rm T}$ scaling analysis can contribute to an understanding of the particle production mechanism at high energies.

\end{document}